\documentclass[pre,twocolumn,showpacs]{revtex4}%
\usepackage{tabularx}
\usepackage{amsmath}
\usepackage{amsfonts}
\usepackage{amssymb}
\usepackage{graphicx}
\usepackage{hyperref}%
\providecommand{\U}[1]{\protect\rule{.1in}{.1in}}
\providecommand{\U}[1]{\protect\rule{.1in}{.1in}}
\begin{document}
\title{The Geography of Scientific Productivity: Scaling in U.S. Computer Science}
\author{Rui Carvalho}
\email{rui.carvalho@ucl.ac.uk}
\affiliation{Centre for Advanced Spatial Analysis, 1-19 Torrington Place, University
College London, WC1E 6BT United Kingdom}
\author{Michael Batty}
\email{mbatty@geog.ucl.ac.uk}
\affiliation{Centre for Advanced Spatial Analysis, 1-19 Torrington Place, University
College London, WC1E 6BT United Kingdom}

\pacs{89.65.-s, 89.75.Da, 89.75.Fb, 89.90.+n}

\begin{abstract}
Here we extract the geographical addresses of authors in the Citeseer database
of computer science papers. We show that the productivity of research centres
in the United States follows a power-law regime, apart from the most
productive centres for which we do not have enough data to reach definite
conclusions. To investigate the spatial distribution of computer science
research centres in the United States, we compute the two-point correlation
function of the spatial point process and show that the observed power-laws do
not disappear even when we change the physical representation from
geographical space to cartogram space. Our work suggests that the effect of
physical location poses a challenge to ongoing efforts to develop realistic
models of scientific productivity. We propose that the introduction of a fine
scale geography may lead to more sophisticated indicators of scientific output.

\end{abstract}
\maketitle

\section{Introduction}

In the last decade, the analysis of mankind's scientific endeavour has become
a rapidly expanding interdisciplinary field. This has been mainly due to the
advent of comprehensive online preprint servers and paper repositories, from
which patterns of productivity and collaboration networks of individual
scientists can be readily ascertained \cite{Shiffrin04}. The vast amount of
available data raises the hope that scientists and policy makers will soon be
able to gain unprecedented insights into the location of research centres and
their productivity. Indeed, little is known today about the influence that
geographical location may have on "invisible colleges" \footnote{An invisible
college is a loose network of researchers who "communicate with each other and
transmit information across the whole field (...) to monitor the rapidly
changing research 'front'." \cite[p35]{Crane72}.} (but see \cite{BattyEPA03,
Borner06}). Conversely, we are only just beginning to uncover how the
historical growth of these "invisible colleges" generates heterogeneities in
the physical location of research centres and, therefore, of the scientists themselves.

Previous investigations of bibliometric data \cite{Egghe90} by physicists have
followed two main directions. On one hand, efforts have focused on
characterizing the topological structure of collaboration networks
\cite{NewmanPNAS01, NewmanPRE01a, NewmanPRE01b, BarabasiPhysicaA02}. On the
other, researchers have used tools of statistical physics to gain insight into
the growth dynamics of scientific outputs \cite{PlerouNature99, Amaral01,
Matia05}. Despite this considerable progress, the relation of collaboration
networks to the productivity of scientists depends on the still poorly
understood fine geographical location of research centres.

Matia \emph{et al.} approached the challenge of characterizing institutional
productivity by analyzing $408$ U.S. institutes for the $11$ year period
$1991-2001$ \cite{Matia05}. They observe a bimodal distribution and conjecture
that this is indicative of a clustering effect of institutes of two different
size classes \cite{Matia05}.

The characterization of spatial structures at large geographical scales has a
long tradition. In $1971$, Glass and Tobler were the first to apply the radial
distribution function (or two-point correlation function, as it is known in
astrophysics \cite{Peacock99}) to the study of cities on a part of the Spanish
plateau \cite{Tobler71, Ripley77}. They choose a $40$ mile square, homogeneous
in town size and density, and apply concepts developed in the study of the
statistical mechanics of equilibrium liquids. Although their analysis does not
detect clustering, we would expect the two-point correlation function to
reveal patterns of concentration and clustering in data whose population sizes
vary over many orders of magnitude.

Recently, Yook \emph{et al.} showed that the nodes of the internet are
embedded on a fractal support driven by the fractal structure of the
population worldwide \cite{Yook02}. This suggests that, in spatial networks
with strong geographical constraints, the nodes may not be distributed
randomly in space \cite{Boccaletti06}, but may be clustered as a function of
population density. Further, Gastner and Newman presented an algorithm based
on physical diffusion to draw density equalizing maps, or cartograms, in which
the sizes of geographic regions appear in proportion to their population or
some other property \cite{Gastner04}. Cartograms give us a tool to probe into
the dependence of one spatial variable (e.g. cancer occurrences) upon another
(e.g. population). In particular, processes which are spatially clustered, but
dependent on population densities, are expected to display random spatial
distributions once the data are transformed by the cartogram \cite{Gastner04,
Gastner06a}.

In order to bring the productivity of research centres and their spatial
interaction patterns under a single roof, we follow a different, but
complementary approach to the ones presented above. Indeed, research centres
are not homogeneously distributed in geographical space and it is likely that
location will impact on their productivity and the structure of collaboration
networks. However, to fully understand the role of location on the production
of science and its networks, one must first characterize the underlying
spatial processes, and this is the road we take here. We therefore investigate
scientific productivity as a function of fine scale geographical location.
Furthermore, to underpin these results, we characterize the spatial point
process generated by the physical location of research centres.

To investigate the role of fine scale geography in the production of science,
one needs to analyze a large dataset. Traditional investigations of
bibliometric data have been carried out by analyzing databases like PubMed,
arXiv.org or Thomson ISI. However, these databases suffer from drawbacks.
Either the data contains only the address of the first (PubMed) or
corresponding author (arXiv.org), or researchers are not uniquely associated
with their addresses (Thomson ISI).

A more promising source of data is the Citeseer digital library, created in
$1998$ as a prototype of Autonomous Citation Indexing \cite{Giles04}. Citeseer
locates computer science articles on the web in Postscript or PDF format and
extracts citations from and to documents \cite{Goodrum01}. Citeseer has made
its metadata available online
\footnote{\href{http://citeseer.ist.psu.edu/oai.html}{http://citeseer.ist.psu.edu/oai.html}
\newline Accessed 22/02/2006.} and the inclusion of an address and affiliation
fields for each author allows a first rigorous analysis into the geography of
a very large bibliometric database.

\section{Spatial Structure}

We studied the Citeseer metadata, which contains $716,772$ records, some of
which are repeated and some of which have authors with empty address fields.
We considered the $N=379,111$ ($52.9\%$) unique papers for which citeseer
identifies all authors and their respective addresses. Out of these $N$ unique
papers, we analyzed the $M=128,348$ ($p_{US}=33.9\%$) papers which have one or
more U.S. authors. Interestingly, $p_{US}$, is in reasonable accordance with
Thomson ISI global indicators, which state that between $1997 $ and $2001$,
the United States output $34.86$ \% of the world's highly cited publications
\cite{King04}.

For each paper, we extracted the $5$--digit ZIP\ code from each author's
address field and geocoded this ZIP into a $(latitude,longitude)$ pair of
coordinates
\footnote{\href{http://www.census.gov/geo/www/tiger/zip1999.html}{http://www.census.gov/geo/www/tiger/zip1999.html}%
}. We identified ZIP\ codes from the address field, by using regular
expressions to match a five-digit code (plus the optional four digit code,
which we ignored) preceded or followed by a U.S. state (or its abbreviation)
or the acronym \emph{USA}. This will leave out addresses like \emph{Roma
00185, Italy} or \emph{Israel 84105}, but will also fail to locate the address
\emph{Physics Department, Northeastern University, Boston MA USA} as it lacks
a ZIP code. We restricted the analysis to the $48$ conterminous U.S. states
plus the District of Columbia.

We identified a total of $116,771$ distinct authors with a U.S. address. Out
of these, $103,928$ ($89\%$) list a single ZIP code in their address, $10,579
$ ($9.1\%$) belong to institutions located in two ZIP codes and $2,264 $
($1.9\%$) \ are located in three or more institutions.

\subsection{Productivity of Research Centres}

To investigate the concept of scaling in publication output of academic
research centres, we computed the probability distribution of total paper
output per ZIP code. We note that ZIP codes were not aggregated. If two
research centres belonging to the same institution have addresses with
distinct ZIP codes, we considered them as distinct centres. This has the
disadvantage of possibly counting more than one research centre per
institution (instead of aggregating both to the same institution). However,
Citeseer covers scientific articles in the field of computer science and it
would be the exception that one institution would have several geographically
separated computer science centres.

Our analysis identified $3,393$ different ZIP codes that matched the
U.S.\ census bureau tables. We implemented a version of fractional counting
\cite{Price81, Egghe90} to compute the productivity of U.S. research centres.
For every paper, we parsed each author's address field and extracted the ZIP
codes therein (there may be more than one ZIP, if the author belongs to more
than one U.S. institution). Each occurrence of a ZIP code in an address field
of a paper increments the productivity of the research centre physically
located at that ZIP code by $1/\phi$, where the normalization factor $\phi$ is
computed as follows. For every address field in the paper being analyzed, we
made $\phi:=\phi+1$ if the address contains no ZIP codes (i.e. it is a
non-U.S. address), or $\phi:=\phi+m$ if the address contains $m\geq1$ ZIP
codes (in which case that specific author will belong to $m$ distinct U.S. institutions).

Identifying research centres by ZIP code has the advantage of simplifying the
data parsing algorithm, which is why we preferred this method to others based
on aggregation by host institution. However, the method is an approximation,
as it cannot distinguish between non-U.S. addresses.

Table \ref{Table1} displays the five most productive ZIP codes and their host
institutions. Interestingly, the two most productive institutions, Carnegie
Mellon University and MIT are also the two most acknowledged entities as shown
by Giles and Councill in a previous study \cite{Giles04}.

\begin{table}[ptb]
\noindent\begin{tabularx}{\linewidth}{|c|c|c|X|}
\hline
Rank&
Zip&
Fractional Count&
Institution\\
\hline
1&
15213&
2343.36&
\raggedright\arraybackslash Carnegie Mellon University\\
2&
02139&
1891.18&
\raggedright\arraybackslash MIT\\
3&
94305&
1512.12&
\raggedright\arraybackslash Stanford University\\
4&
94720&
1496.76&
\raggedright\arraybackslash University of California, Berkeley\\
5&
20742&
1144.70&
\raggedright\arraybackslash University of Maryland, College Park\\
\hline
\end{tabularx}
\caption{Most productive ZIP codes and respective Universities.}%
\label{Table1}%
\end{table}

\begin{figure}
[ptb]
\begin{center}
\includegraphics[
height=5.4613in,
width=3.5475in
]%
{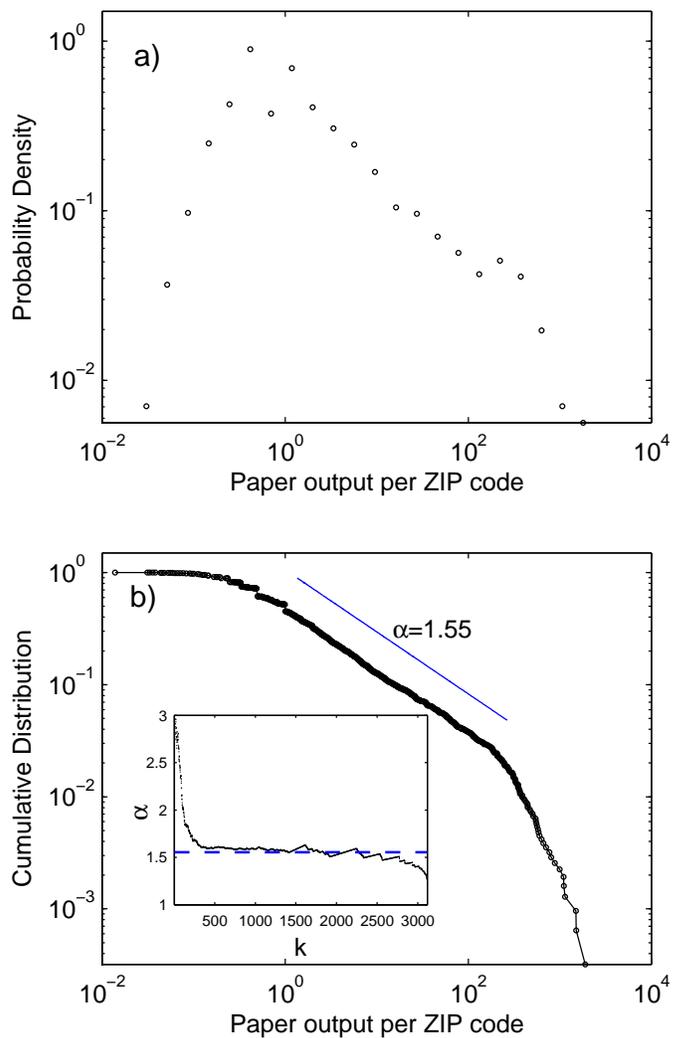}%
\caption{Probability distribution of paper output (fractional counts) per ZIP
code. (a) The probability density is bi-modal and can be approximated by a
power-law regime between the two local maxima. (b) A least squares fit to the
linear region of the cumulative distribution yields $\alpha=1.55$. The inset
shows the Hill plot \cite{Resnick99} as the number of upper order statistics,
$k$, is varied. The match between the plateau on the Hill plot and the least
squares fit (dashed horizontal line), shows that our estimate of $\alpha$ is
appropriate.}%
\label{Fig1}%
\end{center}
\end{figure}

We then asked the question: what is the probability distribution of the
research output of each research centre? To investigate this, we plot the
probability density and cumulative distribution ($P\left[  X>x\right]
=\int_{x}^{\infty}p\left(  y\right)  dy$) in Figure \ref{Fig1}. We found a
bimodal probability distribution of research output by ZIP code (see Figure
\ref{Fig1}a), in agreement with a previous study of the Thomson ISI database
by Matia et. al \cite{Matia05}.

Our results suggest that this probability distribution displays power-law
decay up to the "knee" where the regime changes. Data was insufficient to
determine whether the upper tail of the distribution also decays as a
power-law, albeit with a different exponent. This observation is in apparent
contradiction with the findings of Matia \emph{et al.} who do not find a
power-law regime. The authors\textit{\ }examine the productivity of $408$ U.S.
institutes, whereas our method revealed that papers had been output at $3,393$
U.S. institutes. Therefore, the power-law decay which we observed may be due
to our methodology which included all research institutes in the metadata. On
the other hand, our analysis was limited in scope to the Citeseer database,
whereas Matia \emph{et al.} analyze the Thomson ISI dataset, hence comparisons
with their wider study are necessarily inconclusive. Nevertheless, our results
raise the question of whether power-law decay only appears once one is able to
identify a large percentage of all research institutes.%

\subsection{The Pulling Power of Research Clusters}%

\begin{figure*}[tbp] \centering
{\parbox[b]{7.1754in}{\begin{center}
\includegraphics[
height=4.5645in,
width=7.1754in
]%
{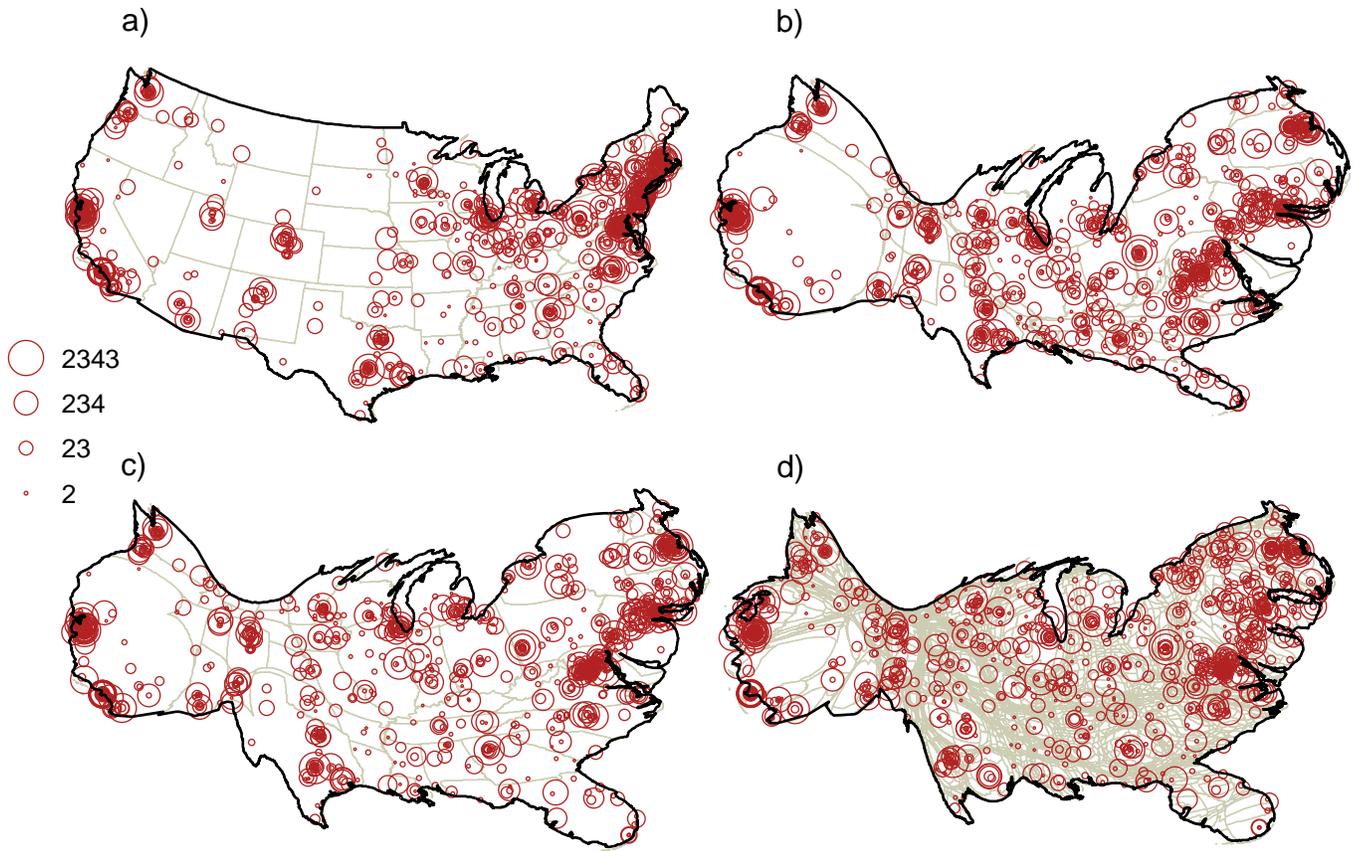}%
\\
{}%
\end{center}}}%
\caption{a) Albers' equal-area
projection of the 48 conterminous states of the US plus the district of
Columbia. Research centres are identified by circles with area proportional
to their productivity on a logarithmic scale. b)-d) Data in a) after a
cartogram transformation with R\&D expenditure by state (b), and population
by state (c) and county (d), respectively. For each panel, we trace the
$14,605$ point border polygon used in the computation of the two-point
correlation function.}\label{Fig2}%
\end{figure*}%

A simple point process in $%
\mathbb{R}
^{2}$ may be considered as a random countable set $X\subset%
\mathbb{R}
^{2}$. The first moment of a point process can be specified by a single
number, the \emph{intensity}, $\rho$, giving the expected number of points per
unit area. The second moment can be specified by Ripley's $K$ function
\cite{Ripley77}, where $\rho K(r)$ is the expected number of points within
distance $r$ of an arbitrary point of the pattern.

The product density%
\begin{equation}
\rho_{2}\left(  \mathbf{x}_{1},\mathbf{x}_{2}\right)  dA\left(  \mathbf{x}%
_{1}\right)  dA\left(  \mathbf{x}_{2}\right)  =\rho^{2}g\left(  r\right)
dA\left(  \mathbf{x}_{1}\right)  dA\left(  \mathbf{x}_{2}\right)
\label{TwoPointCF}%
\end{equation}
describes the probability to find a point in the area element $dA\left(
\mathbf{x}_{1}\right)  $ and another point in $dA\left(  \mathbf{x}%
_{2}\right)  $, at the distance $r=\left\vert \mathbf{x}_{1}-\mathbf{x}%
_{2}\right\vert $, and $g\left(  r\right)  $ is the two-point correlation
function. Ripley's $K$ function \ is related to $g\left(  r\right)  $ by
\cite{Stoyan2000}
\begin{equation}
K\left(  r\right)  =2\pi\int g\left(  r\right)  rdr\label{RipleysK}%
\end{equation}
In other words, $g\left(  r\right)  $ is the density of $K\left(  r\right)  $
with respect to the radial measure $rdr$. The benchmark of complete randomness
is the spatial Poisson process, for which $g\left(  r\right)  =1$ and
$K(r)=\pi r^{2}$, the area of the search region for the points. Values larger
than this indicate clustering on that distance scale, and smaller values
indicate regularity.

The two-point correlation function can be estimated from $N$ data points
$\boldsymbol{x}\in D$ inside a sample window $\mathcal{W}$ by
\cite{Kerscher00}:%
\begin{equation}
g(r)=\frac{\left\vert \mathcal{W}\right\vert }{N\left(  N-1\right)  }%
\sum_{x\in D}\sum_{y\in D}\frac{\Phi_{r}\left(  \boldsymbol{x},\boldsymbol{y}%
\right)  }{2\pi r\Delta}\omega\left(  \boldsymbol{x},\boldsymbol{y}\right)
\label{2 point correlation}%
\end{equation}
where $2\pi r\Delta$ is the area of the annulus centred at $\boldsymbol{x}$
with radius $r$ and thickness $\Delta$. Here $\left\vert \mathcal{W}%
\right\vert $ is the area of the sample window, and the sum is restricted to
pairs of different points $\boldsymbol{x}\neq\boldsymbol{y}$. The function
$\Phi_{r}$ is symmetric in its argument and $\Phi_{r}\left(  \boldsymbol{x}%
,\boldsymbol{y}\right)  =\left[  r\leq d\left(  \boldsymbol{x},\boldsymbol{y}%
\right)  \leq r+\Delta\right]  $, where $d\left(  \boldsymbol{x}%
,\boldsymbol{y}\right)  $ is the Euclidean distance between the two points and
the condition in brackets equals $1$ when true and $0$ otherwise.

The function $\omega\left(  \boldsymbol{x},\boldsymbol{y}\right)  $ accounts
for a bounded $\mathcal{W}$ by weighting points where the annulus intersects
the edges of $\mathcal{W}$. There are a number of edge-corrections available,
but that of Ripley \cite{Ripley76} has a long tradition both in human
geography \cite{Ripley77} and physics \cite{Kerscher00}:
\begin{equation}
\omega\left(  \boldsymbol{x},\boldsymbol{y}\right)  =\frac{2\pi r}%
{\digamma\left(  \partial\mathcal{B}_{r}\left(  \boldsymbol{x}\right)
\cap\mathcal{W}\right)  }\label{RipleyWeights}%
\end{equation}
where $\digamma\left(  \partial\mathcal{B}_{r}\left(  \boldsymbol{x}\right)
\cap\mathcal{W}\right)  $ is the fraction of the perimeter of the circle
$\mathcal{B}_{r}\left(  \boldsymbol{x}\right)  $ with radius $r=\left\vert
\boldsymbol{x}-\boldsymbol{y}\right\vert $ around $\boldsymbol{x}$ inside
$\mathcal{W}$ --e.g. $\digamma\left(  \partial\mathcal{B}_{r}\left(
\boldsymbol{x}\right)  \cap\mathcal{W}\right)  =\pi r$ if only half of the
annulus falls inside $\mathcal{W}$. Note that $\omega\left(  \boldsymbol{x}%
,\boldsymbol{y}\right)  =1$ iff $\partial\mathcal{B}_{r}\left(  \boldsymbol{x}%
\right)  \subset\mathcal{W}$, in which case the summand in
(\ref{2 point correlation}) is simply the sum of $\Phi_{r}\left(
\boldsymbol{x},\boldsymbol{y}\right)  $ weighted by the area of the annulus
centred at $\boldsymbol{x}$ with radius $r$ and thickness $\Delta$. If
$\partial\mathcal{B}_{r}\left(  \boldsymbol{x}\right)  \cap\overline
{\mathcal{W}}\neq\varnothing,$ that is the circle $\mathcal{B}_{r}\left(
\boldsymbol{x}\right)  $ is only partially in the sample window $\mathcal{W}$,
then $\Phi_{r}\left(  \boldsymbol{x},\boldsymbol{y}\right)  $ is weighted by
the area of the fraction of the annulus which is inside $\mathcal{W}$.

Of special physical interest is whether the two-point correlation is
scale-invariant. A scale-invariant $g\left(  r\right)  $ is an indicator of a
fractal distribution of research centres, and is expected in critical
phenomena \cite{Kerscher2000}.

To investigate the presence of power-law decay in the two-point correlation
function we selected the $1,046$ research centres (ZIP codes) which had a
total fractional count of two papers or more. We chose this productivity
threshold for two main reasons. A first factor was to consider only research
centres which can be clearly identified as active. Second, the computation of
the two-point correlation function requires reasonable computer resources as
$\mathcal{W}$ is a fine boundary of the United States --in our case, a polygon
with $14,605$ points.

Next, we projected the U.S.\ map and the $(latitude,longitude)$ pairs of the
research centres with the Albers' equal area projection \cite{Robinson95}%
\footnote{\href{http://www.census.gov/geo/www/cob/}{http://www.census.gov/geo/www/cob/}%
} and computed the two-point correlation function, $g\left(  r\right)  $, of
the resulting point process.

To investigate whether the decay of $g\left(  r\right)  $ is a function of the
distribution of R\&D funding or population, we applied several cartogram
transformations to the base map and the points: first, we computed the
cartogram projection using U.S. R\&D funding expenditure, by state, for the
year $2001$ \cite[table B-17]{NSF05}; second we computed the cartogram with
U.S. population, by state and county, from the $2000$ census
\footnote{\href{http://www.census.gov/popest/datasets.html}{http://www.census.gov/popest/datasets.html}%
}. The points representing the research centres were transformed accordingly
to each cartogram. Figure \ref{Fig2}a) shows the Albers' equal area projection
and each centre is represented by a circle with area proportional to the
number of papers output on a logarithmic scale. Figures \ref{Fig2}b)-d) show
the cartograms with R\&D expenditure by state, and population by state and
county, respectively. It is obvious from these maps that as the cartogram
transformation uses finer spatial scales (e.g. from U.S. states to counties),
the points become more homogeneously distributed spatially.

\begin{figure}
[h]
\begin{center}
\includegraphics[
height=2.7614in,
width=3.4445in
]%
{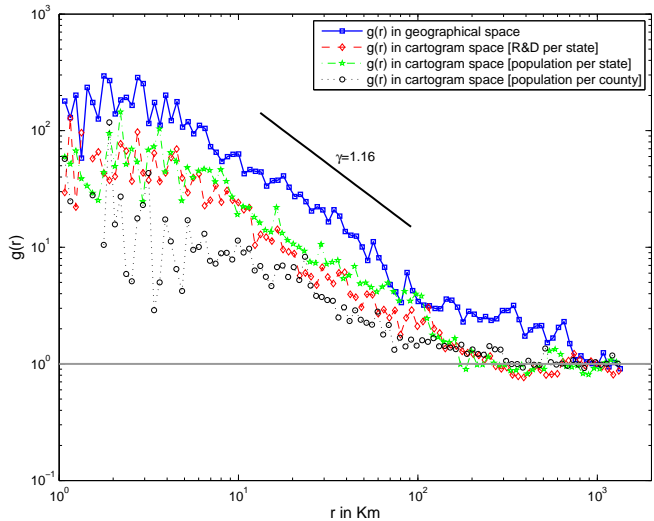}%
\caption{Variation of the two-point correlation function with distance (Km).
In blue, $g\left(  r\right)  $ computed from projections of the border and the
points with the Albers' equal area projection. In\ red, green and black,
$g\left(  r\right)  $ computed from further transforming the data by the
cartogram projection with R\&D per state, population per state and county,
respectively. The horizontal line at $g=1$ is the expected value of $g\left(
r\right)  $ for a Poisson process.}%
\label{Fig3}%
\end{center}
\end{figure}

The two-point correlation function computed for the projected data (see Figure
\ref{Fig2}a)) is plot in Figure \ref{Fig3}, where we observe a power-law decay
$g\left(  r\right)  \sim r^{-\gamma}$ with $\gamma\simeq1.16$. Next we asked
the following question: can the power-law decay of $g\left(  r\right)  $ be
explained by a clustering of research centres in areas where research funding
or population is higher? To answer this question, we computed $g\left(
r\right)  $ for the same point process, but now using the data transformed by
the cartograms with R\&D expenditure by state (Figure \ref{Fig2}b)),
population by state (Figure \ref{Fig2}c)), and population by county (Figure
\ref{Fig2}d)). Our results showed that the power-law decay was still present
after the cartogram projections, although as the transformation was performed
at finer spatial scales, $g\left(  r\right)  $ approached the expected value
for a Poisson process, $g\left(  r\right)  =1$, at shorter distances.%

\section{Discussion}

Considerable advances have been made over the past few years in understanding
the structure of scientific production and its networks. Along this road,
physicists have computed a number of quantities to characterize networks of
scientific collaborations, mainly by analyzing data from online preprint
servers and repositories. However, these studies have not addressed the impact
of fine scale physical location on the statistical characterization of the
scientific enterprise and it networks. Here we have presented a detailed study
of the productivity of research centres in U.S. computer science (identified
by ZIP codes) and characterized the pattern of spatial concentration which
these centres display.

A first important conclusion of our study is that the productivity of U.S.
research centres in computer science was highly skewed. A surprising result of
our study was the power-law decay of the probability distribution of research
output for some orders of magnitude. A second important conclusion is that the
physical location of research centres in the U.S. formed a fractal set, which
was not completely destroyed by population or research funding patterns.

Although we consider our results to be promising, there are still several
caveats. First our conclusions are clearly only valid for the U.S.
\cite{Moed99, Matia05} and even from the Citeseer database, which we consider
is the best currently available for such analysis, there are problems of
missing and inaccurate data which we are not able to quantify. Nevertheless,
our results are consistent with those from the burgeoning geography of
information technology which suggests in qualitative fashion, that such
technologies are correlated with population but also have their own dynamic
\cite{Zook05, Dodge06}. In this sense, our result that the scaling inherent in
the geographical distribution of paper production in U.S. computer science is
still present once the geography has been normalized with respect to the
distribution of population and R\&D expenditures, implies processes that are
endogenous to the dynamics of research \cite{Amaral01}.

In summary, the method introduced in this paper could serve as a starting
point for an investigation of the role of the fine scale physical location of
research centres in the production of science. Our study focused on U.S.
computer science but further analyses should be possible as preprint server
repositories make more elaborate metadata available. And such developments may
lead to a better understanding of the role of physical location not just in
science, but for a much wider class of complex spatial systems.

\begin{acknowledgments}
We wish to thank Michael Gastner (SFI) for help with the code to generate
cartograms and Isaac Councill (Penn State) for help with the Citeseer
database. This research was supported by the Engineering and Physical Sciences
Research Council under grant EP/C513703/1.
\end{acknowledgments}

\end{document}